\documentclass{aa}
\usepackage{txfonts}
\usepackage{graphicx}
\usepackage{natbib}
\bibpunct{(}{)}{;}{a}{}{,}

\begin{document}

\def\vx{\vec{x}}
\def\vr{\vec{r}}
\def\vk{\vec{k}}
\def\vab{\varepsilon_B}
\def\vabi{\varepsilon_{B_p}}
\def\ud{{\rm d}}
\def\obs{{\rm obs}}
\def\true{{\rm true}}
\def\min{{\rm min}} 
\def\max{{\rm max}}

\def\del#1{{}}

\title{A Bayesian view on Faraday rotation maps - \\Seeing the magnetic
power spectra in galaxy clusters}
\titlerunning{Bayesian view on $RM$ maps}
\author{Corina Vogt \and Torsten A. En{\ss}lin} 
\authorrunning{Corina Vogt \and Torsten A. En{\ss}lin}
\institute{Max-Planck-Institut f\"{u}r
Astrophysik, Karl-Schwarzschild-Str.1, Postfach 1317, 85741 Garching,
Germany} 
 
\date{Submitted / Accepted} 

\abstract { 

We present a Bayesian maximum likelihood analysis of Faraday rotation
measure ($RM$) maps of extended radio sources to determine magnetic
field power spectra in clusters of galaxies. Using this approach, it is
possible to determine the uncertainties in the measurements. We apply
this approach to the $RM$ map of Hydra A and derive the power spectrum
of the cluster magnetic field. For Hydra A, we measure a spectral index
of $-5/3$ over at least one order of magnitude implying Kolmogorov type
turbulence. We find a dominant scale $\sim 3$ kpc on which the magnetic
power is concentrated, since the magnetic autocorrelation length is
$\lambda_B = 3 \pm 0.5$ kpc. Furthermore, we investigate the influences
of the assumption about the sampling volume (described by a window
function) on the magnetic power spectrum. The central magnetic field
strength was determined to be $\sim$ 7 $\pm\,2\,\mu$G for the most
likely geometries.

\keywords{ Radiation mechanism: non-thermal -- Galaxies: active --
Intergalactic medium -- Galaxies: cluster: general -- Radio continuum: 
general } }

\maketitle

\section{Introduction\label{sec:intro}}
The intra-cluster medium is magnetised. Direct evidence for
cluster-wide magnetic fields are the large-scale diffuse radio sources
of synchrotron origin. There is growing evidence that these fields are
of the order of $\sim \mu$G and are ordered on kiloparsec scales
\citep[see e.g. recent reviews][]{2002ARA&A..40..319C,
2002RvMP...74..775W, 2004astro.ph.10182G}.
  
One method to investigate magnetic field structure and strength is the
detection of the Faraday rotation effect. This effect is observed
whenever linearly polarised radio emission passes through a magnetised
medium. A linearly polarised wave can be described by two circularly
polarised waves. Their motion along magnetic field lines in a plasma
introduces a phase difference between the two waves resulting in a
rotation of the plane of polarisation. If the Faraday active medium is
external to the source of the polarised emission, one expects the
change in polarisation angle to be proportional to the squared
wavelength. The proportionality factor is called the rotation measure
($RM$). This quantity can be evaluated in terms of the line of sight
integral over the product of the electron density and the magnetic
field component along the line of sight.

Observed $RM$ maps of extended extragalactic radio sources are
especially valuable in order to study the intra-cluster magnetic
fields. Simple analytical approaches based on the patchy structure of
the $RM$ maps to measure the characteristic length scale of the
magnetic fields, which are necessary to translate $RM$ values into
field strength, result in magnetic field strength of $\sim$ 5 $\mu$G up
to $\sim$ 30 $\mu$G for cooling flow clusters, e.g. Cygnus A
\citep{1987ApJ...316..611D}, Hydra A \citep{1993ApJ...416..554T}, A1795
\citep{1993AJ....105..778G}, 3C295 \citep{2001MNRAS.324..842A}. The
same arguments have lead to estimates of a cluster magnetic field
strength of 2...8 $\mu$G for non-cooling flow clusters, e.g. Coma
\citep{1995A&A...302..680F}, A119 \citep{1999A&A...344..472F}, 3C129
\citep{2001MNRAS.326....2T}, A2634 \& A400 \citep{2002ApJ...567..202E}.

Observations of a polarised radio point source sample seen through a
  cluster atmosphere were presented by
  \citet{1991ApJ...379...80K}. They detected an $RM$ broadening towards
  the cluster centre implying a magnetic field strength of 1
  $\mu$G. More recently, \citet{2001ApJ...547L.111C} analysed a
  statistical sample of 16 cluster sources against a control
  sample. They also detect a broadening of the $RM$ distribution for
  sources towards the cluster centre. They find a cluster magnetic
  field strength of \hbox{4...8 $\mu$G}.

These high magnetic field values derived using $RM$ methods seem to be
  in contrast to the lower magnetic field values of 0.1...0.3 $\mu$G
  estimates from Inverse Compton (IC) measurements which are possible
  for clusters with observed diffuse radio haloes
  \citep{1987ApJ...320..139R, 1994ApJ...429..554R, 1999ApJ...511L..21R,
  1998PASJ...50..389H, 2000ApJ...534L...7F, 2001ApJ...552L..97F,
  2004ApJ...602L..73F, 1998A&A...330...90E}. Cosmic microwave
  background photons are expected to inverse Compton scatter off of the
  relativistic electrons thereby emitting non-thermal X-ray
  emission. Upper limits on this non-thermal X-ray emission together
  with the radio observations of the synchrotron radiation which is
  emitted by the relativistic electron population can then be used to
  set lower limits on the average magnetic field strength.

There is an order of magnitude difference between the field strength
  derived for these methods. Several arguments can be given to
  reconcile the different results. First, except for a very small
  number of clusters (including the Coma cluster), at best one of the
  methods could be applied, so that the difference could be a
  difference between clusters. Second, the Faraday rotation method
  measures a volume-averaged magnetic field weighted by the thermal
  electron density whereas the inverse Compton results give
  volume-averaged field strengths which are weighted with the
  relativistic electron distribution. Since the relativistic electron
  population is easily diminished in regions with strong magnetic
  fields due to the enhanced synchrotron cooling, the inverse Compton
  method is expected to provide smaller estimates. Thus, a medium
  that is inhomogeneously magnetised on small scales compared to the
  observational spatial resolution might possibly solve the
  contradiction \citep{1999A&A...344..409E}. Furthermore, since the
  observed IC flux could originate from other sources, it is an upper
  limit. Hence, the IC measurements give only lower limits on the
  magnetic field strength. For a more detailed discussion, we refer to
  \citet{2002ARA&A..40..319C, 2004astro.ph.10182G}.

\citet{2003A&A...401..835E} proposed a method to determine the magnetic
power spectra by Fourier transforming $RM$ maps. Based on these
considerations, \citet{2003A&A...412..373V} applied this method and
determined the magnetic power spectrum of three clusters (Abell 400,
Abell 2634 and Hydra~A) from $RM$ maps of radio sources located in
these clusters. Furthermore, they determined field strengths of $\sim
12\,\,\mu$G for the cooling flow cluster Hydra~A, $3 \,\, \mu$G and $6
\,\, \mu$G for the non-cooling flow clusters Abell 2634 and Abell 400,
respectively. Their analysis revealed spectral slopes of the power
spectra with spectral indices $ -2.0\ldots-1.6$. However, it was
realised that using the proposed analysis, it is difficult to reliably
determine differential quantities such as spectral indices due to the
complicated shapes of the emission regions used which lead to a
redistribution of magnetic power within the spectra.

Recently, \citet{2004A&A...424..429M} proposed a numerical method to
determine the magnetic power spectrum in clusters. They infer the
magnetic field strength and structure by comparing simulations of $RM$
maps as caused by multi-scale magnetic fields with the observed
polarisation properties of extended cluster radio sources such as radio
galaxies and haloes. They argue that field strengths derived in the
literature using analytical expressions have been overestimated by a
factor of $\sim$ 2.

In order to determine a power spectrum from observational data, maximum
likelihood estimators are widely used in astronomy. These methods and
algorithms have been greatly improved, especially by the Cosmic
Microwave Background (CMB) analysis which tackles the problem of
determining the power spectrum from large CMB
maps. \citet{1998ApJ...495..564K} proposed such an estimator to
determine the power spectrum of a primordial magnetic field from the
distribution of $RM$ measurements of distant radio galaxies.

Based on the initial idea of \citet{1998ApJ...495..564K}, the methods
developed by the CMB community
\citep[especially][]{1998PhRvD..57.2117B} and our understanding of the
magnetic power spectrum of cluster gas \citep{2003A&A...401..835E}, we
derive here an Bayesian maximum likelihood approach to calculate the
magnetic power spectrum of cluster gas given observed Faraday rotation
maps of extended extragalactic radio sources. The power spectrum
enables us also to determine characteristic field length scales and
strength. After testing our method on artificially generated $RM$ maps
with known power spectra, we apply our analysis to a Faraday rotation
map of Hydra~A. The data were kindly provided by Greg Taylor. In
addition, this method allows us to determine the uncertainties of our
measurement and, thus, we are able to give errors on the calculated
quantities. Based on these calculations, we investigate the nature of
turbulence of the magnetised gas.

This paper is structured as follows. In Sect.~\ref{sec:theory}, a
method employing a maximum likelihood estimator as suggested by
\citet{1998PhRvD..57.2117B} to determine the magnetic power spectrum
from $RM$ maps is introduced. Special requirements for the analysis of
$RM$ maps with such a method are discussed. In Sect.~\ref{sec:test}, we
apply our maximum likelihood estimator to generated $RM$ maps with
known power spectra to test our algorithm. In Sect.~\ref{sec:app}, the
application of our method to data of Hydra~A is described. In
Sect.~\ref{sec:discussion}, the derived power spectra are presented and
the results are discussed.  In Sect.~\ref{sec:conclusion}, conclusions
are drawn.

We assume a Hubble constant of H$_{0} = 70$ km s$^{-1}$ Mpc$^{-1}$,
$\Omega_{m} = 0.3$ and $\Omega_{\Lambda} = 0.7$ in a flat universe. All
equations follow the notation of \citet{2003A&A...401..835E}.

\section{Maximum likelihood analysis\label{sec:theory}}

\subsection{The covariance matrix $C_{RM}$\label{sec:crm}}
One of the most commonly used methods of Bayesian statistics is the
maximum likelihood method. The likelihood function for a model
characterised by $p$ parameters $a_p$ is equivalent to the probability
of the data $\vec{\Delta}$ given a particular set of $a_p$ and can be
expressed in the case of (near) Gaussian statistics of $\vec{\Delta}$
as
\begin{equation}
\label{eq:likely}
\mathcal{L}_{\vec{\Delta}}(a_p) = \frac{1}{(2\pi)^{n/2}|C|^{1/2}}\cdot
\exp\left(-\frac{1}{2}\vec{\Delta}^{T}C ^{-1}\vec{\Delta}\right),
\end{equation}
where $|C|$ indicates the determinant of a matrix, $\Delta_i =
RM_i$ are the actual observed data, $n$ indicates the number of
observationally independent points and $C = C(a_p)$ is the
covariance matrix. This covariance matrix can be defined as
\begin{equation}
C_{ij}(a_p) = \langle \Delta_i^{obs}\Delta_j^{obs} \rangle = \langle
RM_i^{obs}\,RM_j^{obs} \rangle,
\end{equation}
where the brackets $\langle \rangle$ denote the expectation value and,
thus, $C_{ij}(a_p)$ describes our expectation based on the proposed
model characterised by a particular set of $a_p$s. Now, the likelihood
function $\mathcal{L}_{\vec{\Delta}}(a_p)$ has to be maximised for the
parameters $a_p$. Although the magnetic fields might be non-Gaussian,
the $RM$ should be close to Gaussian due to the central limit
theorem. Observationally, $RM$ distributions are known to be close to
Gaussian \citep[e.g.][]{1993ApJ...416..554T, 1999A&A...344..472F,
1999A&A...341...29F, 2001MNRAS.326....2T}.

Ideally, the covariance matrix is the sum of a signal and a noise
matrix term which results if the errors are uncorrelated to true
values. Writing $RM^{obs} = RM^{true} + \delta RM$ results in
\begin{eqnarray}
C_{ij}(a_p) & = & \langle RM_i^{true} RM_j^{true} \rangle + \langle
\delta RM_i \, \delta RM_j \rangle \nonumber \\ 
& = & C_{RM}(\vx_{\perp i},\, \vx_{\perp j}) + \langle \delta RM_i \,
\delta RM_j \rangle
\end{eqnarray}
where $\vx_{\perp i}$ is the displacement of point $i$ from the
$z$-axis and $\langle \delta RM_i \, \delta RM_j \rangle$ indicates the
expectation for the uncertainty in our measurement. Unfortunately,
while in the discussion of the power spectrum measurements of CMB
experiments the noise term is extremely carefully studied, for our
discussion this is not the case and goes beyond the scope of the
paper. Thus, we will neglect this term. However,
\citet{1995MNRAS.273..877J} discuss uncertainties involved in the data
reduction process to gain a model for $\langle \delta RM_i\, \delta
RM_j \rangle$.

Since we are interested in the magnetic power spectrum, we have to
find an expression for the covariance matrix $C_{ij}(a_p) =
C_{RM}(\vx_{\perp i},\,\vx_{\perp j})$ which can be identified as the
$RM$ autocorrelation $\langle RM(\vx_{\perp i})\,RM(\vx_{\perp j})
\rangle$. This has then to be related to the magnetic power spectra.

The observable in any Faraday experiment is the rotation measure
\textit{RM}. For a line of sight parallel to the $z$-axis and
displaced by $\vx_{\perp}$ from it, the $RM$ arising from polarised
emission passing from the source $z_s(\vx_{\perp})$ through a
magnetised medium to the observer located at infinity is expressed by
\begin{equation}
RM(\vx_{\perp})= a_0 \int_{z_s(\vx_{\perp})}^{\infty} \!\!\! \ud \vx\,
n_{{\rm e}}(\vx) \, B_z (\vx),
\end{equation}
where $a_0 = e^3/(2\pi m_e^2c^4)$, $\vx = (\vx_\perp, z)$, $n_e(\vx)$
is the electron density and $B_z(\vx)$ is the magnetic field component
along the line of sight.

In the following, we will assume that the magnetic fields in galaxy
clusters are isotropically distributed throughout the Faraday
screen. If one samples such a field distribution over a large enough
volume they can be treated as statistically homogeneous and
statistically isotropic. Therefore, any statistical average over a
field quantity will not be influenced by the geometry or the exact
location of the volume sampled. Following \citet{2003A&A...401..835E},
we can define the elements of the $RM$ covariance matrix using the
$RM$ autocorrelation function $C_{RM}(\vx_{\perp i}, \vx_{\perp j}) =
\left< RM(\vx_{\perp i})RM(\vx_{\perp j}) \right>$ and introduce a
window function $f(\vx)$ which describes the properties of the
sampling volume
\begin{equation}
\label{eq:correl}
C_{RM}(\vx_{\perp}, \vx'_{\perp}) = \tilde{a_0}^2 \!\!\! \int_{z_s}
^\infty \!\!\!\!\!\! \ud z \int_{z'_s} ^ \infty \!\!\!\!\!\! \ud
z' f(\vx)f(\vx')\left< B_z(\vx_{\perp}, z) B_z(\vx'_{\perp}, z')
\right>,
\end{equation}
where $\tilde{a_0} = a_0n_{e0}$, the central electron density is
$n_{e0}$ and the window function is defined by
\begin{equation}
\label{eq:window}
f(\vx) = \mathbf{1}_{\{\vx_{\perp} \in \Omega\} }\,\mathbf{1}_{\{z
\geq z_{\rm s}(\vx_{\perp})\}} \, \,g(\vx) \,n_e(\vx)/n_{e0},
\end{equation}
where $\mathbf{1}_{\{condition\}}$ is equal to unity if the condition
is true and zero if not and $\Omega$ defines the region for which
$RM$s were actually measured. The electron density distribution
$n_e(\vx)$ is chosen with respect to a reference point $\vx_{ref}$
(usually the cluster centre) such that $n_{e0} = n_e(\vx_{ref})$,
e.g. the central density, and $B_0 = \langle \vec{B}^2 (\vx_{ref})
\rangle ^{1/2}$. The dimensionless average magnetic field profile
$g(\vx) = \langle \vec{B} ^2 (\vx) \rangle ^{1 / 2} / \vec{B} _{0}$ is
assumed to scale with the density profile such that $g(\vx) =
(n_e(\vx)/n_{e0})^{\alpha_{B}}$.

Setting $\vx' = \vx + \vr$ and assuming that the correlation length of
the magnetic field is much smaller than characteristic changes in
the electron density distribution, we can separate the two integrals
in Eq.~(\ref{eq:correl}). Furthermore, we can introduce the magnetic
field autocorrelation tensor \hbox{$M_{ij} = \langle B_i(\vx) \cdot
B_j(\vx+\vr) \rangle$} \citep[see e.g.][]{ 1999PhRvL..83.2957S,
2003A&A...401..835E}. Taking this into account, the
$RM$ autocorrelation function can be described by
\begin{equation}
\label{eq:sep_int}
C_{RM}(\vx_{\perp}, \vx_{\perp} + \vr_{\perp}) = \tilde{a_0}^2
\int_{z_s} ^\infty \!\!\!\! \ud z \, f(\vx)f(\vx+\vr) \int_{(z'_s - z)
\to -\infty} ^ \infty \!\!\!\!\!\!\!\!\!\!\!\!\! \ud r_z M_{zz}(\vr)
\end{equation}
Here, the approximation $(z'_s - z) \to -\infty$ is valid for Faraday
screens which are much thicker than the magnetic autocorrelation
length. This will turn out to be the case in the application at hand.

The Fourier transformed $zz$-component of the autocorrelation tensor
$M_{zz}(\vec{k})$ can be expressed by the Fourier transformed scalar
magnetic autocorrelation function $w(k) = \sum_i M_{ii}(k)$ and a $k$
dependent term (see Eq.~(31) in \citet{2003A&A...401..835E}) leading to
\begin{equation}
\label{eq:mzz_r}
M_{zz}(\vr) = \frac{1}{2\pi^3} \int ^\infty _{-\infty} \!\!\! \ud ^3k
\,\frac{w(k)}{2}\,\left( 1 - \frac{k_z^2}{k^2} \right) \, {\rm
e}^{-i\vk \vr}
\end{equation}
Furthermore, the one dimensional magnetic energy power spectrum
$\vab(k)$ can be expressed in terms of the magnetic
autocorrelation function $w(k)$ such that
\begin{equation}
\label{eq:wk_ebk}
\vab(k)\, \ud k = \frac{k^2w(k)}{2\,(2\pi)^3}\, \ud k.
\end{equation}

As stated in \citet{2003A&A...401..835E}, the $k_z = 0$ - plane of
$M_{zz}(\vec{k})$ is all that is required to reconstruct the magnetic
autocorrelation function $w(k)$. Thus, inserting Eq.~(\ref{eq:mzz_r})
into Eq.~(\ref{eq:sep_int}) and using Eq.~(\ref{eq:wk_ebk}) leads to 
\begin{eqnarray}
C_{RM}(\vx_{\perp}, \vx_{\perp} + \vr_{\perp}) & = & 4\pi^2
\tilde{a_0}^2 \int_{z_s} ^\infty \!\!\!\!\! \ud z\, f(\vx)f(\vx+\vr)
\times \nonumber \\
& & \int_{-\infty} ^ \infty \!\!\!\!\! \ud k\, \vab(k)
\frac{J_0(kr_{\perp})}{k},
\end{eqnarray}
where $J_0(kr_{\perp})$ is the 0th Bessel function. This equation
gives an expression for the $RM$-autocorrelation function in terms of
the magnetic power spectra of the Faraday-producing medium.  

Since the magnetic power spectrum is the interesting function, we
parametrise $\vab(k) = \sum_p \vabi \mathbf{1}_{\{ k \, \in \, [k_p,
k_{p+1}] \}}$, where $\vabi$ is constant in the interval $[k_p,
k_{p+1}]$, leading to
\begin{equation}
\label{eq:cfinal}
C_{RM}(\varepsilon_{B_p}) = 4\pi^2 \tilde{a_0}^2 \int_{z_s} ^\infty
\!\!\!\!\!\!dz\, f(\vx)f(\vx+\vr) \sum_p\! \varepsilon_{B_p}
\int_{k_p} ^ {k_{p+1}} \!\!\!\!\!\!\!\! dk\,
\frac{J_0(kr_{\perp})}{k},
\end{equation}
where the $\varepsilon_{B_p}$ are to be understood as the model
parameter $a_P$ for which the likelihood function
$\mathcal{L}_{\vec{\Delta)}}(a_p)$ has to be maximised given the Faraday
data $\vec{\Delta}$.

\subsection{Evaluation of the likelihood function\label{sec:likely}}
In order to maximise the likelihood function,
\citet{1998PhRvD..57.2117B} approximate the likelihood function as a
Gaussian of the parameters in regions close to the maximum $\vec{a} =
\{ a \}_{\max}$, where $\{ a \}_{\max}$ is the set of model parameters
which maximise the likelihood function. In this case, one can perform
a Taylor expansion of $\ln\mathcal{L}_{\vec{\Delta}}(\vec{a}+\delta
\vec{a})$ about $a_p$ and truncates at the second order in $\delta
a_p$ without making a large error.
\begin{eqnarray}
\ln \mathcal{L}_{\vec{\Delta}}(\vec{a}+\delta \vec{a}) & = & \ln
\mathcal{L}_{\vec{\Delta}}(\vec{a}) + \sum_p \frac{\partial
\ln\mathcal{L}_{\vec{\Delta}}(\vec{a})}{\partial a_{p}} \delta a_p +
\nonumber \\
& & \frac{1}{2} \sum_{pp'} \frac{\partial^2
\ln\mathcal{L}_{\vec{\Delta}}(\vec{a})} {\partial a_p\, \partial a_{p'}}
\delta a_p \delta_{p'}
\end{eqnarray}

With this approximation, one can directly solve for the $\delta a_p$
that maximise the likelihood function $\mathcal{L}$
\begin{equation}
\label{eq:delta}
\delta a_p = - \sum_{p'} \left( \frac{\partial^2
\ln\mathcal{L}_{\vec{\Delta}}(\vec{a})} {\partial a_p\, \partial a_{p'}}
\right)^{-1}\, \frac{\partial
\ln\mathcal{L}_{\vec{\Delta}}(\vec{a})}{\partial a_{p'}},
\end{equation}
where the first derivative is given by
\begin{equation}
\label{eq:first}
\frac{\partial \ln\mathcal{L}_{\vec{\Delta}}(\vec{a})}{\partial
a_{p'}} = \frac{1}{2} \mathrm{Tr} \left[ \left( \vec{\Delta}
\vec{\Delta}^T - C \right) \left( C^{-1} \frac{\partial C}{\partial
a_{p'}} C^{-1} \right) \right]
\end{equation} 
and the second derivative is expressed by 
\begin{eqnarray}
\label{eq:second}
 \mathcal{F}^{(a)} _{pp'} & = & - \left( \frac{\partial^2
\ln\mathcal{L}_{\vec{\Delta}}(\vec{a})} {\partial a_p\, \partial
a_{p'}} \right) = \mathrm{Tr} \left[ \left( \vec{\Delta}
\vec{\Delta}^T - C \right) \left( C^{-1} \frac{\partial C}{\partial
a_{p}} C^{-1}\frac{\partial C}{\partial a_{p'}} C^{-1}
\right. \right. \nonumber \\
& & \left. \left.  - \frac{1}{2} C^{-1}\frac{\partial^2 C}{\partial a_p
\partial a_{p'}} C^{-1} \right) \right] + \frac{1}{2} \mathrm{Tr}
\left( C^{-1} \frac{\partial C}{\partial a_{p}} C^{-1} \frac{\partial
C}{\partial a_{p'}} \right),
\end{eqnarray}
where Tr indicates the trace of a matrix. The second derivative is
called the curvature matrix. If the covariance matrix is linear in the
parameter $a_p$ then the second derivatives of the covariance matrix
$\partial^2 C/(\partial a_p \partial a_{p'})$ vanish. Note that for
the calculation of the $\delta a_p$, the inverse curvature matrix
$(\mathcal{F}^{(a)} _{pp'})^{-1}$ has to be calculated. The diagonal
terms of the inverse curvature matrix $(\mathcal{F}_{pp} ^{(a)})^{-1}$
can be regarded as the errors $\sigma^2 _{a_p}$ to the parameters
$a_p$.

A suitable iterative algorithm to determine the power spectra would be
to start with an initial guess of a parameter set $a_p$. Using this
initial guess, the $\delta a_p$s have to be calculated using
Eq.~(\ref{eq:delta}). If the $\delta a_p$s are not sufficiently close
to zero, a new parameter set $a' _p = a_p + \delta a_p$ is used and
again the $\delta a' _p$ are calculated and so on. This process can be
stopped when $\delta a_p / \sigma_{a_p} \le \epsilon$, where
$\epsilon$ describes the required accuracy.

\subsection{Binning and rebinning}
In our parametrisation of the model given by Eq.~(\ref{eq:cfinal}) the
bin size, i.e.~the size of the interval $[k_p, k_{p+1}]$, is
important. Since we are measuring the power spectrum, we chose equal
bins on a logarithmic scale as the initial binning. However, if the
bins are too small then the cross correlation between two bins could be
very high and the two bins cannot be regarded as independent
anymore. Furthermore, the errors might be very large, and could be one
order of magnitude larger than the actual values. In order to avoid
such situations, it is preferable to chose either fewer bins or to
rebin by adding two bins together. Note that this oversampling is not a
real problem, since the model parameter covariance matrix takes care of
the redundancy between data points. However, for computational
efficiency and for a better display of the data, a smaller set of
mostly independent data points is preferable.

To find a criterion for rebinning, an expression for the
cross correlation of two parameter $a_p$ and $a_{p'}$ can be defined
by
\begin{equation}
\label{eq:cross}
\delta_{pp'} = \frac{\langle \sigma_p \sigma_{p'} \rangle}{\langle
\sigma_p\rangle \, \langle \sigma_{p'} \rangle} = \frac{\mathcal{F}^{-1}
_{pp'}}{\sqrt{\mathcal{F}^{-1} _{pp} \mathcal{F}^{-1} _{p'p'}}},
\end{equation} 
where the full range, $-1 \le \delta_{pp'} \le 1$, is possible but
usually the correlation will be negative, indicating
anti-correlation. Our criterion for rebinning is to require that if
the absolute value of the cross-correlation $| \delta_{pp'} |$ is
larger than $\delta_{pp'} ^{\rm max}$ for two bins $p$ and $p'$ then
these two bins are added together in such a way that the magnetic
energy $\sum_p \varepsilon_{B_p}* \Delta k_{p}$ is conserved.

After rebinning the algorithm again starts to iterate and finds the
maximum with the new binning. This is done as long as the
cross-correlation of two bins is larger than required.

\subsection{The algorithm}

As a first guess for a set of model parameter $\vabi$, we used the
results from a Fourier analysis of the original $RM$ map employing the
algorithms as described in \citet{2003A&A...412..373V}. However, we
also employed as first guess $\vabi$ a simple power law $\vabi \propto
k_i^{\alpha}$, where $\alpha$ is the spectral index. The results and the
shape of the power spectrum did not change.

If not stated otherwise, an iteration is stopped when $\epsilon <
0.01$, i.e. the change in parameter $\vabi$ is smaller than 1\% of
the error in the parameter $\vabi$ itself. Once the iteration
converges to a final set of model parameters the cross-correlation
between the bins is checked and if necessary, the algorithm will start
a new iteration after rebinning. Throughout the rest of the paper, we
require a $| \delta_{pp'} | < 0.5$ for $p \neq p'$.

Once the power spectra in terms of $\vab(k) = \sum _p \vabi
\mathbf{1}_{\{[k_p, k_{p+1}]\}}$ is determined, we can calculate the
magnetic energy density $\vab$ by integration of the power spectrum
\begin{equation}
\vab(a_p) = \int _0 ^{\infty} \ud k\, \vab(k) = \sum_p \vabi \Delta
k_p,
\end{equation}
where $\Delta k_p = k_{p+1} - k_p$ is the binsize.

Also $\lambda_B$ and $\lambda_{RM}$ are accessible by integration of
the power spectrum \citep{2003A&A...401..835E}. 
\begin{eqnarray}
\lambda_B & = & \pi \frac{\int_0 ^{\infty} \ud k \, \vab(k)/k}{\int _0
^{\infty} \ud k \, \vab(k)} = \pi \frac{\sum_p \vabi
\ln(k_{p+1}/k_{p})} {\sum_p \vabi \Delta k_p} \\ 
\lambda_{RM} & = & 2 \frac{\int_0 ^{\infty} \ud k \,
\vab(k)/k^2}{\int_0 ^{\infty} \ud k \, \vab(k)/k} = 2 \frac{\sum_p
\vabi \left( 1/k_{p} - 1/k_{p+1} \right)}{\sum_p \vabi
\ln(k_{p+1}/k_{p})}.
\end{eqnarray}

Since the method allows to calculate errors $\sigma_{\vabi}$, one can
also determine errors for these integrated quantities. However, the
cross-correlations $\delta_{pp'}$ which are non-zero as already
mentioned, have to be taken into account. The probability distribution
$P(\vec{a})$ of a parameter can often be described by a Gaussian
\begin{equation}
\label{eq:prob}
P(\vec{a}) \sim e^{-\frac{1}{2} \delta \vec{a} ^T X^{-1} \delta \vec{a}},
\end{equation}
where $X$ is the covariance matrix of the parameters, $\delta \vec{a}
= \vec{a} - \vec{a}_{{\rm peak}}$, $\vec{a}=\{a\}_{\max}$ is the
determined maximum value for the probability distribution and
$\vec{a}_{{\rm peak}}$ is the actual maximum of the probability
function. The standard deviation is defined as
\begin{equation}
\label{eq:deltaeb}
\langle \delta \vab^2 \rangle = \langle (\vab(a)-\vab)^2 \rangle =
\int\, \ud^n a\, P(a)\,(\vab(a) - \vab)^2.
\end{equation}
Assuming that $P(\vec{a})$ follows a Gaussian distribution (as done above in
Eq.~(\ref{eq:prob})) and using that $\vab(a)$ is linear in the $a_p =
\vabi$ then Eq.~(\ref{eq:deltaeb}) becomes
\begin{eqnarray}
\langle \delta \vab^2 \rangle & = & \int \ud^n a\, P(a) \, \left[
\delta a \frac{\partial \vab}{\partial a_p} \right]^2\\
& = & \int \ud^n a\, P(a) \, \sum_p \delta a_p \, \frac{\partial
\vab}{\partial a_p} \, \sum_{p'}
\delta a_{p'} \, \frac{\partial \vab}{\partial a_{p'}}.
\end{eqnarray}

Rearranging this equation and realising that the partial derivatives
are independent of the $a_p$ since $\vab$ is linear in the $a_p$s this
leads to
\begin{equation}
\langle \delta \vab^2 \rangle = \sum_{pp'} \frac{\partial \vab}{\partial
a_p}\, \frac{\partial \vab}{\partial a_{p'}} \int \ud^n a \, P(a) \, \delta
a_p \, \delta a_{p'}
\end{equation}
and finally using Eq.~(\ref{eq:cross}) 
\begin{equation}
\langle \delta \vab^2 \rangle = \sum_{pp'} \frac{\partial
\vab}{\partial a_p}\, \frac{\partial \vab}{\partial a_{p'}} \,\langle
\sigma_{p} \sigma_{p'} \rangle,
\end{equation}
where $\langle \sigma_{p} \, \sigma_{p'} \rangle = \mathcal{F}_{pp'} ^{-1}$

A similar argumentation can be applied to the error derivation for the
correlation lengths $\lambda_{RM}$ and $\lambda_B$, although the
correlation length are not linear in the coefficients $a_p$. If one
uses the partial derivatives at the determined maximum, one still is
able to approximately separate them from the integral. This leads to
the following expressions for their errors
\begin{equation}
\langle \delta \lambda_B^2 \rangle \approx \sum_{pp'} \frac{\partial
\lambda_B}{\partial a_p} \bigg\arrowvert_{a _p ^{\max}} \frac{\partial
\lambda_B}{\partial a_{p'}} \bigg\arrowvert_{a _{p'} ^{\max}} \langle
\sigma_{p} \sigma_{p'} \rangle
\end{equation}
and
\begin{equation}
\langle \delta \lambda_{RM}^2 \rangle \approx \sum_{pp'} \frac{\partial
\lambda_{RM}}{\partial a_p} \bigg\arrowvert_{a _p ^{\max}}
\frac{\partial \lambda_{RM}}{\partial a_{p'}} \bigg\arrowvert_{a _{p'}
^{\max}} \langle \sigma_{p} \sigma_{p'} \rangle.
\end{equation}

\section{Testing the algorithm\label{sec:test}}
\begin{figure*}[htb]
\resizebox{\hsize}{!}{\includegraphics{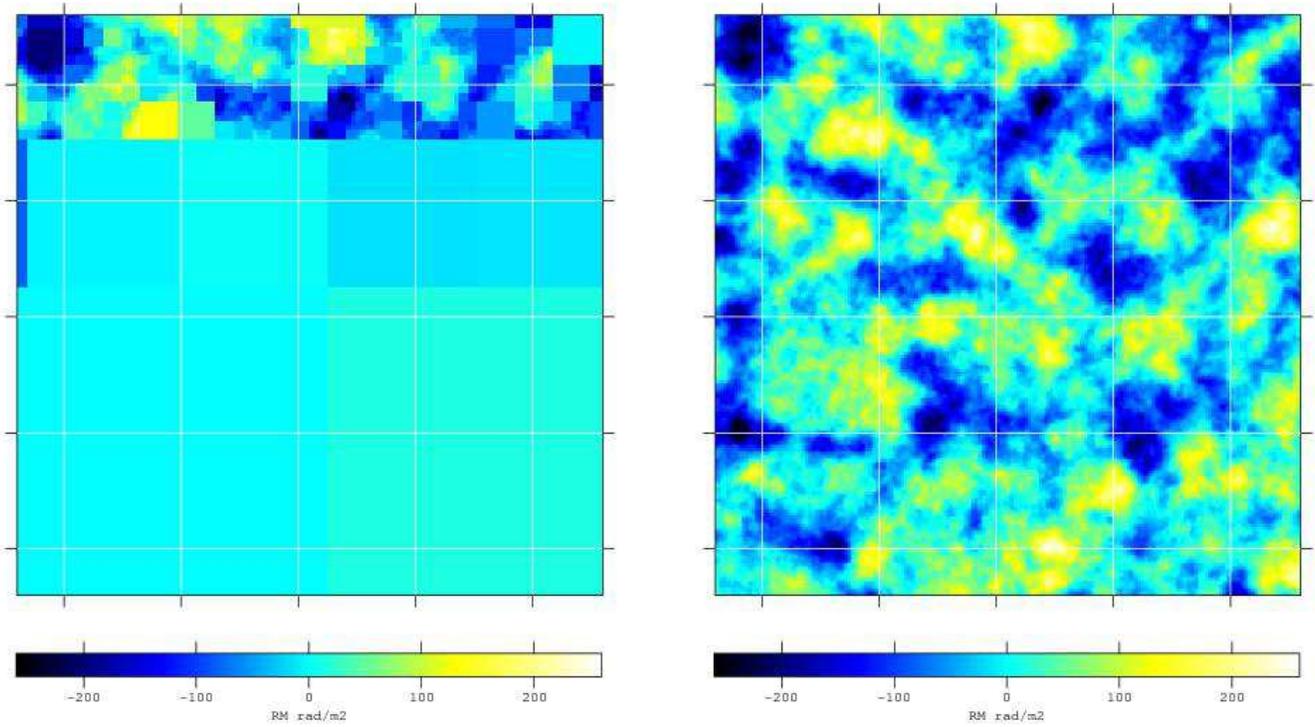}}
\caption[]{\label{fig:testrm} Right panel, a small part ($37 \times 37$
kpc) of a typical realisation of a $RM$ map which is produced by a
Kolmogorov-like magnetic field power spectrum for $k \geq k_c = 0.8$
kpc$^{-1}$ and a magnetic field strength of 5 $\mu$G. Left panel, the
$RM$ data used for the data matrix $\Delta_i$ is shown where we
averaged arbitrary neighbouring points in order to reduce the number of
independent points in a similar way as done later with the
observational data.}
\end{figure*}

In order to test our algorithm, we applied our maximum likelihood
estimator to generated $RM$ maps with a known magnetic power spectrum
$\vab(k)$. \citet{2003A&A...401..835E} give a prescription (their
Eq.~(37)) for the relation between the amplitude of $RM$,
$|\hat{RM}(k_{\perp})|^2$, and the magnetic power spectrum in Fourier
space
\begin{equation}
\varepsilon_B^{\obs}(k) = \frac{k^2}{a_1\,A_{\Omega}(2\pi)^4}
\int_{0}^{2\pi} \ud \phi \,\, |\hat{RM}(\vk_{\perp})|^2
\end{equation}
or
\begin{equation}
\label{eq:rmk}
|\hat{RM}(k_{\perp})|^2 = \frac{a_1\,A_{\Omega}(2\pi)^3}{k^2}
 \varepsilon_B^{\obs}(k),
\end{equation}
where $A_{\Omega}$ is the area $\Omega$ for which $RM$'s are actually
measured and $a_1 = a_0^2\,n_{{\rm e0}}^2\,L$, where $L$ is the
characteristic depth of the Faraday screen.

As the Faraday screen, we assumed a box with sides being 150 kpc long
and a depth of $L = 300$ kpc. For simplicity, we assumed a uniform
electron density profile with a density of $n_{{\rm e0}} = 0.001$
cm$^{-3}$. For the magnetic field power spectra, we used
\begin{equation}
\label{eq:k0norm}
\varepsilon_B^{\obs}(k) = \left\{ 
\begin{array}{ll} 
\frac{\vab}{k_0^{1-\alpha} \, k_c^{2+\alpha}} \, k^2 & \forall k \leq
k_c \\
\frac{\vab}{k_0}\left( \frac{k}{k_0} \right)^{-5/3} & \forall k
\geq k_c 
\end{array} \right. .
\end{equation}
where the spectral index which was set to mimic Kolmogorov turbulence
with energy injection at $k = k_{c}$, and
\begin{equation} 
\vab = \frac{\langle B^2 \rangle}{8\pi} = \int _{0} ^{k_{\max}}
\!\!\! \ud k \, \vab ^{\obs}(k),
\end{equation}
where $k_{\max} = \pi/\Delta r$ is determined by the pixel size
($\Delta r$) of the $RM$ map used. The latter equation combined with
Eq.~(\ref{eq:k0norm}) gives the normalisation $k_0$ in such a way that
the integration over the accessible power spectrum will result in a
magnetic field strength of $B$ for which we used 5 $\mu$G. We used a
$k_{c} = 0.8$ kpc$^{-1}$.

In order to generate a $RM$ map with the magnetic power spectrum
$\vab(k)$ for the chosen Faraday screen, we filled the real and
imaginary part of the Fourier space independently with Gaussian
deviates. Then these values were multiplied by the appropriate values
given by Eq.~(\ref{eq:rmk}) corresponding to their place in
$k$-space. As a last step, an inverse Fourier transformation was
performed. A typical realisation of such a generated $RM$ map is shown
in Fig.~\ref{fig:testrm}.

\begin{figure}[htb]
\resizebox{\hsize}{!}{\includegraphics{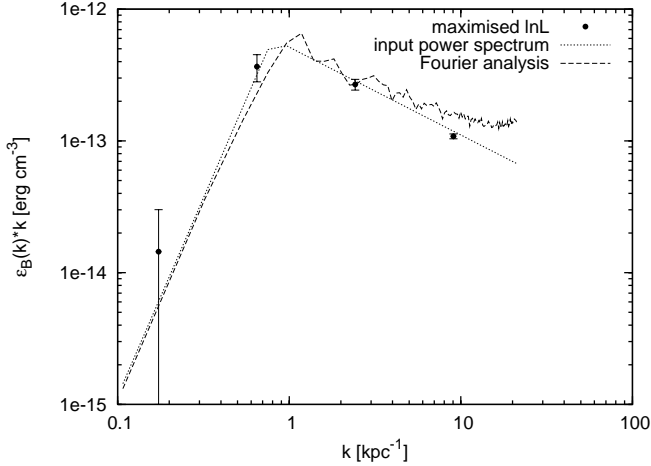}}
\caption[]{\label{fig:test} Power spectra for a simulated $RM$ map as
shown in Fig.~\ref{fig:testrm}. The input power spectrum is shown in
comparison to the one found by the Fourier analysis as described in
\citet{2003A&A...401..835E} and the one which was derived by our
maximum likelihood estimator. One can see the good agreement within one
$\sigma$ between input power spectrum and the power spectrum derived by
the maximum likelihood method.}
\end{figure}

For the analysis of the resulting $RM$ map only a small part of the
initial map was used in order to reproduce the influence of the limited
emission region of a radio source. We applied the Fourier analysis as
described in \citet{2003A&A...401..835E} to this part. The resulting
power spectrum is shown in Fig.~\ref{fig:test} as a dashed line in
comparison with the input power spectrum as a dotted line.

The maximum likelihood method is numerically limited by computational
power since it involves matrix multiplication and inversion, where the
latter is a $N^3$ process. Thus, not all points of the many which are
defined in our maps can be used. However, it is desirable to use as
much information as possible from the original map. Therefore we chose
to randomly average neighbouring points with a scheme which let to a
map with spatially inhomogeneously resolved cells. The resulting map is
highly resolved on top and lowest on the bottom with some random
deviations which make it similar to the error weighting of the observed
data. We used $N$ = 1500 independent points for the analysis. In the
left panel of Fig.~\ref{fig:testrm}, the averaged $RM$ map which was
used for the test is shown.

As a first guess for the maximum likelihood estimation, we used the
power spectra derived by the Fourier analysis. The resulting power
spectrum is shown as filled circles with 1-$\sigma$ error bars in
Fig.~\ref{fig:test}. The input power spectrum and the power spectrum
derived by the maximum likelihood estimator agree well within the one
$\sigma$ level. Integration over this power spectrum results in a field
strength of \hbox{$(4.7 \pm 0.3) \mu$G} in agreement with the input
magnetic field strength of \hbox{$5 \mu$G}.

\begin{figure*}[hbt]
\resizebox{\hsize}{!}{\includegraphics{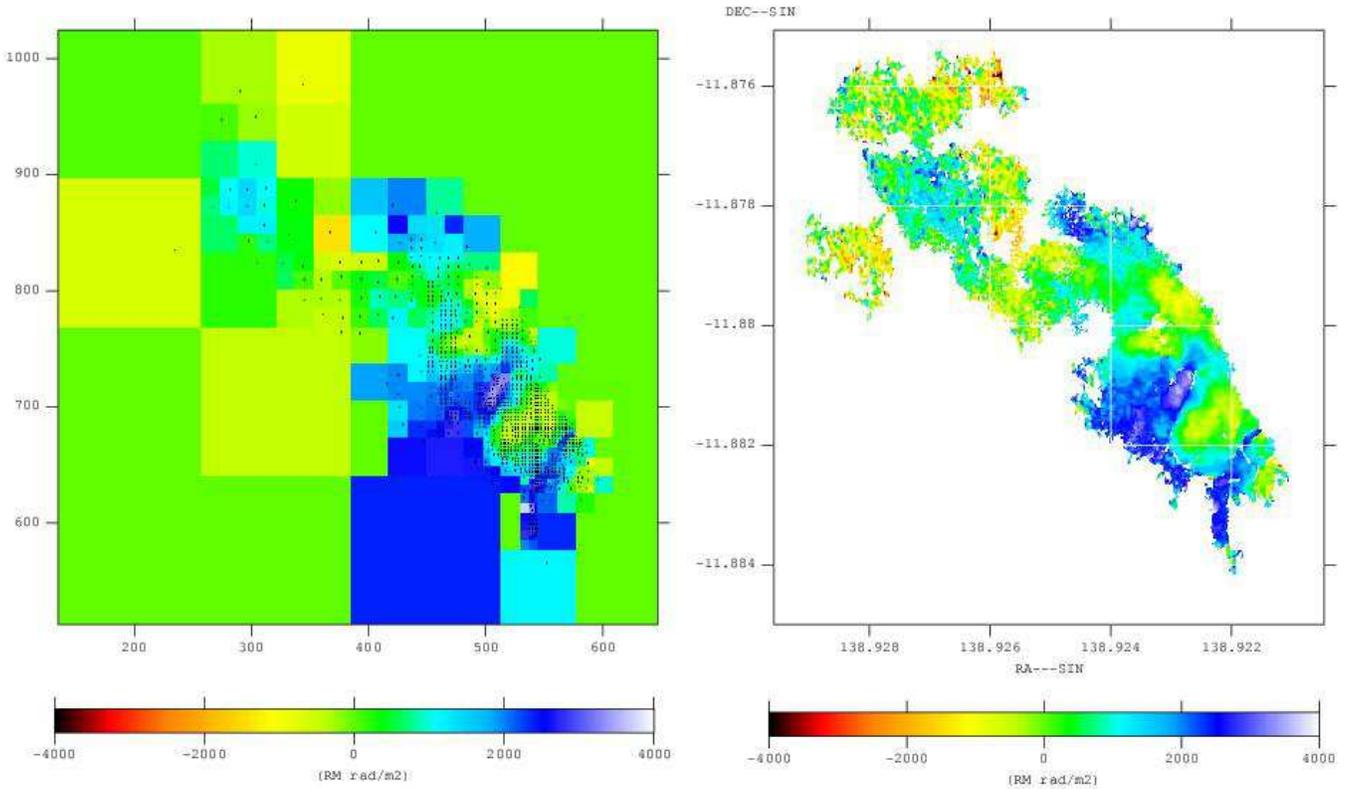}}
\caption[]{\label{fig:rmav} The final $RM$ map from the north lobe of
Hydra~A which was analysed with the maximum likelihood estimator; left:
error weighted map. The dots indicate the coordinates which
correspond to the appropriate error weighted $RM$ value, which resulted
from averaging over the indicated area; right: original
\textit{Pacman} map. Note that the small scale noise for the diffuse
part of the lobe is averaged out and only the large scale information
carried by this region is maintained.}
\end{figure*}

\section{Application to Hydra~A\label{sec:app}}
\subsection{The data $\vec{\Delta}$ \label{sec:data}}

We applied this maximum likelihood estimator introduced and tested in
the last sections to the Faraday rotation map of the north lobe of the
radio source Hydra~A \citep{1993ApJ...416..554T}. The data were kindly
provided by Greg Taylor.

For this purpose, we used a high fidelity $RM$ map presented in
\citet{2004astro.ph..1216V} which was generated by the newly developed
algorithm \textit{Pacman} \citep{2004astro.ph..1214D} using the
original polarisation data. \textit{Pacman} also provides error maps
$\sigma_i$ by error propagation of the instrumental uncertainties of
polarisation angles. The \textit{Pacman} map which was used is shown
in the right panel of Fig.~\ref{fig:rmav}.

For the same reasons as mentioned in Sect.~\ref{sec:test}, we averaged
the data. An appropriate averaging procedure using error weighting was
applied such that
\begin{equation}
\overline{RM}_i = \frac{\sum_j {RM_j}/{\sigma^2_j}} {\sum_{j}
{1}/{\sigma_j^2}},
\end{equation} 
and the error calculates as
\begin{equation}
\sigma ^2 _{\overline{RM}_i} = \frac{\sum_j \left( {1}/
{\sigma^2_j} \right) } { \left( \sum_{j} {1}/{\sigma_j^2}\right)^2 } =
\frac{1}{\sum_{j} {1}/{\sigma_j^2}}.
\end{equation}
Here, the sum goes over the set of old pixels $\{ j \}$ which form the
new pixels $\{ i \}$. The corresponding pixel coordinates $\{ i \}$
were also determined by applying an error weighting scheme
\begin{equation}
\overline{x}_i = \frac{\sum_j {x_j}/{\sigma^2 _j}}{\sum_j
1/\sigma^2_j} \;\;\mbox{and}\;\;
\overline{y}_i = \frac{\sum_j {y_j}/{\sigma^2 _j}}{\sum_j
1/\sigma^2_j}.
\end{equation}

The analysed $RM$ map was determined by a gridding procedure. The
original $RM$ map was divided into four equally sized cells. In each
of these the original data were averaged as described above. Then the
cell with the smallest error was chosen and again divided into four
equally sized cells and the original data contained in the so-determined
cell were averaged. The last step was repeated until the number of
cells reached a defined value $N$. We decided to use $N = 1500$. This
is partly due to the limitation of computational power but also partly
because of the desired suppression of small-scale noise by a strong
averaging of the noisy regions.

The final $RM$ map which was analysed is shown in
Fig.~\ref{fig:rmav}. The most noisy regions in Hydra~A are located in
the coarsely resolved northernmost part of the lobe. We
chose not to resolve this region any further but to keep the
large-scale information which is carried by this region.

\subsection{The window function\label{sec:window}}
As mentioned in Sect.~\ref{sec:crm}, the window function describes the
sampling volume and, thus, we have to find a suitable description for
it based on Eq.~\ref{eq:window}. Hydra A (or 3C218) is located at a
redshift of 0.0538 \citep{1991trcb.book.....D}. For the derivation of
the electron density profile parameter, we relied on the work by
\citet{1999ApJ...517..627M} done for ROSAT PSPC data while using the
deprojection of X-ray surface brightness profiles as described in the
Appendix A of \citet{2004A&A...413...17P}. Since Hydra A is known to
exhibit a strong cooling flow as observed in the X-ray studies, we
assumed a double $\beta$-profile
\footnote {
defined as
\begin{math}
n_{{\rm e}}(r) = [n_{{\rm e1}}^2 (0)(1+(r/r_{{\rm
c1}})^2)^{-3\beta}+n_{{\rm e2}}^2 (0)
(1+(r/r_{{\rm c2}})^2)^{-3\beta}]^{1/2}.
\end{math}
} and used for the inner profile $n_{{\rm e1}}(0) = 0.056$ cm$^{-3}$
and $r_{c1} = 0.53$ arcmin; for the outer profile we used $n_{{\rm
e2}}(0) = 0.0063$ cm$^{-3}$ and $r_{c2} = 2.7$ arcmin and we applied a
$\beta = 0.77$.

\begin{figure*}[hbt]
\resizebox{\hsize}{!}{\includegraphics[angle=-90]{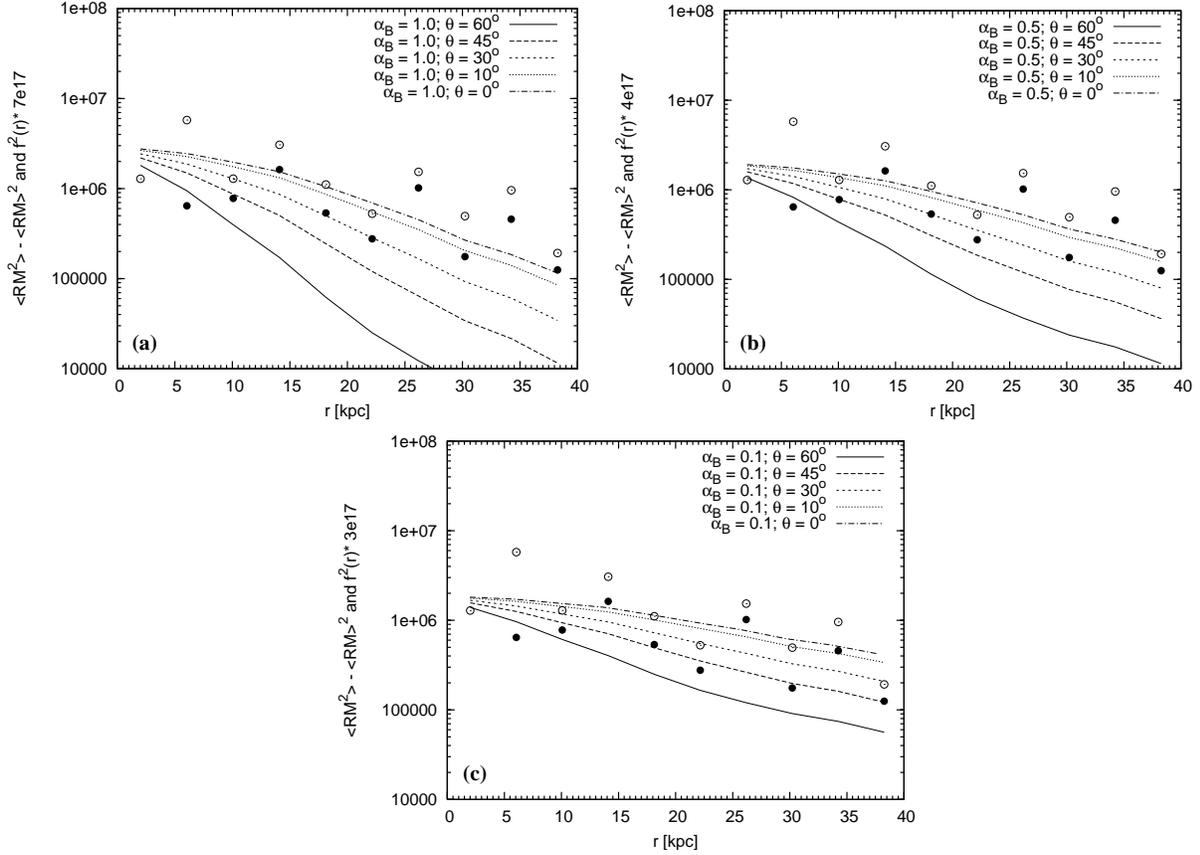}}
\vspace{-1.0cm}
\caption[]{\label{fig:radial} The comparison of the integrated
squared window function $f^2(r)$ (lines) with the $RM$ dispersion
function $\langle RM^2(r) \rangle$ (open circles) and $\langle RM^2
\rangle - \langle RM(r) \rangle^2$ (filled circles). Different models
for the window function were assumed. In (a) $\alpha_B = 1.0$, in (b)
$\alpha_B = 0.5$ and in (c) $\alpha_B = 0.1$ were used, where the
inclination angle $\theta$ of the source was varied. It can be seen
that models for the window function with $\alpha_B = 0.1\ldots0.5$ and
$\theta = 10\degr \ldots 50\degr$ match the shape of the
dispersion function very well.}
\end{figure*}

Assuming this electron density profile to be accurately determined,
there are two other parameters which enter in the window function. The
first one is related to the source geometry. For Hydra~A, a clear
depolarisation asymmetry between the two lobes is observed, known as
the Laing-Garrington effect \citep{1988Natur.331..147G,
1988Natur.331..149L} suggesting that the source is tilted from the
$xy$-plane \citep{1993ApJ...416..554T}. In fact, the north lobe points
towards the observer. In order to take this into account, we
introduced an angle $\theta$ which describes the angle between the
source and the $xy$-plane such that the north lobe points towards the
observer. \citet{1993ApJ...416..554T} determine an inclination angle
of $\theta = 45\degr$.

The other parameter is related to the global magnetic field
distribution which is assumed to scale with the electron density
profile $B(r) \propto n_{{\rm e}}(r) ^ {\alpha_B}$. In a scenario in
which an originally statistically homogeneously magnetic energy
density gets adiabatically compressed, one expects $\alpha_B =
2/3$. If the ratio of magnetic and thermal pressure is constant
throughout the cluster then $\alpha_B = 0.5$. However, $\alpha_B$
might have any other value. \citet{2001A&A...378..777D} determined an
$\alpha_B=0.9$ for the outer regions of the cluster Abell 119.

In order to constrain the applicable ranges of these quantities, one
can compare the integrated squared window function with the $RM$
dispersion function $\langle RM(r_{\perp})^2 \rangle$ of the $RM$ map
used since
\begin{equation}
\langle RM^2 (r_{\perp}) \rangle \propto \int _{-\infty} ^{\infty}
\ud z\, f^2(r_{\perp}, z),
\end{equation}
as stated by Eq.~(24) of \citet{2003A&A...401..835E}. Therefore, we
compared the shape of the two functions. The result is shown in
Fig.~\ref{fig:radial}. For the window function, we used three
different $\alpha_B =0.1, 0.5, 1.0$ and for each of these, five
different inclination angles $\theta = 0\degr, 10\degr, 30\degr,
45\degr$ and $60\degr$ were employed, although the $\theta = 0\degr$
is not very likely considering the observational evidence of the
Laing-Garrington effect as observed in Hydra~A by
\citet{1993ApJ...416..554T}. The different results are plotted as
lines of different style in Fig.~\ref{fig:radial}. The filled and open
dots represent the $RM$ dispersion function. The solid circles
indicate the binned $\langle RM^2 \rangle$ function. The open circles
represent the binned $\langle RM^2 \rangle - \langle RM \rangle^2$
function, which is cleaned of any foreground $RM$ signals.

From Fig.~\ref{fig:radial}, it can be seen that models with $\alpha_B
= 1.0$ or $\theta > 50\degr$ are not able to recover the shape of
the $RM$ dispersion function and, thus, we expect $\alpha_B < 1.0$ and
$\theta < 50\degr$ to be more likely.

\section{Results and discussion\label{sec:discussion}}  
\begin{figure}[htb]
\resizebox{\hsize}{!}{\includegraphics{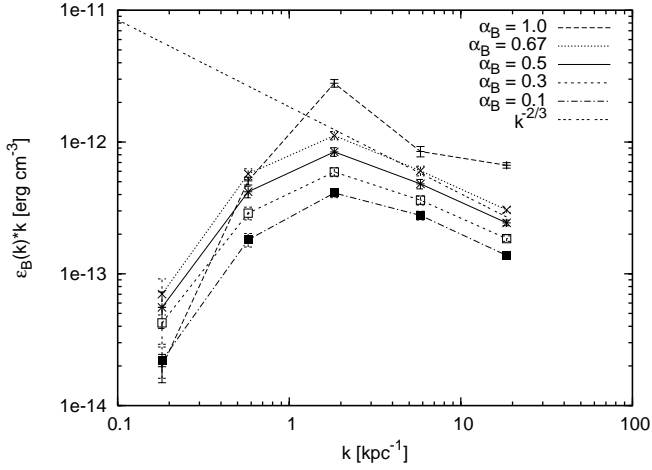}}
\caption[]{\label{fig:power_alpha} Power spectra for $N = 1500$ and
$n_l = 5$. Different exponents $\alpha_B$ in the relation $B(r) \sim
n_e(r)^{\alpha_B}$ of the window function were used. The inclination
angle of the source was chosen to be $\theta = 45\degr$.}
\end{figure}

Based on the described treatment of the data and the description of
the window function, first we calculated power spectra for various
scaling exponents $\alpha_B$ while keeping the inclination angle at
$\theta = 45\degr$. For this investigation, we used as the number of
bins $n_l = 5$ which proved to be sufficient. For these calculations,
we used $\epsilon < 0.1$. The resulting power spectra are plotted in
Fig.~\ref{fig:power_alpha}.

In Fig.~\ref{fig:power_alpha}, one can see that the power spectrum
derived for $\alpha_B = 1.0$ has a completely different shape whereas
the other power spectra show only slight deviation from each other and
are vertically displaced, implying different normalisation factors,
i.e. central magnetic field strengths which increase with increasing
$\alpha_B$. The straight dashed line which is also plotted in
Fig.~\ref{fig:power_alpha} indicates a Kolmogorov-like power spectrum
being equal to $5/3$ in our prescription. The power spectra follow this
slope over at least one order of magnitude.

\begin{figure}[htb]
\resizebox{\hsize}{!}{\includegraphics{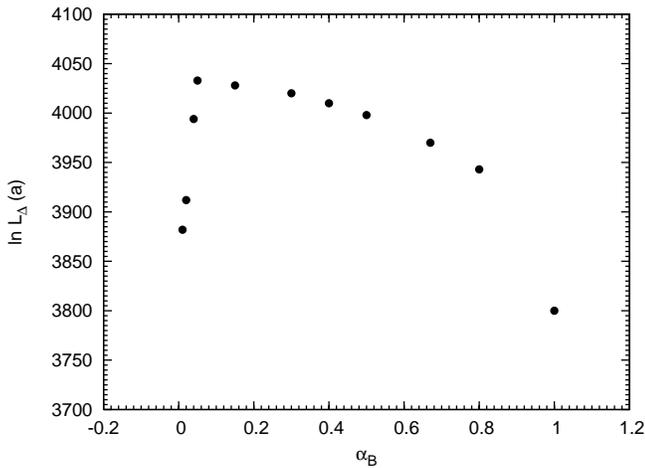}}
\caption[]{\label{fig:alpha_lnL} The log likelihood $\ln \mathcal
{L}_{\vec{\Delta}} (\vec{a})$ of various power spectra assuming
different $\alpha_B$ while using a constant inclination angle $\theta =
45\degr$. $\alpha_B = 0.1\ldots0.8$ are in the plateau of maximum
likelihood. The sudden decrease for $\alpha_B < 0.1$ in the likelihood
might be due to non-Gaussian effects becoming too strong.}
\end{figure}

In Sect.~\ref{sec:window}, we were not able to distinguish between the
various scenarios for $\alpha_B$ although we found that an $\alpha_B =
1$ does not properly reproduce the measured $RM$ dispersion. However,
the likelihood function offers the possibility to calculate the actual
probability of a set of parameters given the data (see
Eq.~(\ref{eq:likely})). Thus, we calculated the log likelihood $\ln
{\mathcal L}_{\vec{\Delta}}(\vec{a})$ value for various power spectra
derived for the different window functions varying in the scaling
exponent $\alpha_B$ and assuming the inclination angle of the source
to be for all geometries $\theta = 45\degr$. In
Fig.~\ref{fig:alpha_lnL}, the log likelihood is shown as a function of
the scaling parameter $\alpha_B$ used.

As can be seen from Fig.~\ref{fig:alpha_lnL}, there is a plateau of
most likely scaling exponents $\alpha_B$ ranging from 0.1 to 0.8. An
$\alpha_B = 1$ seems to be very unlikely for our model as already
deduced in Sect.~\ref{sec:window}. The sudden decrease for $\alpha_B <
0.1$ might be due to non-Gaussian effects. The magnetic field strength
derived for this plateau region ranges from 9 $\mu$G to 5 $\mu$G. The
correlation length of the magnetic field $\lambda_B$ was determined to
range between $2.5$ kpc and $3.0$ kpc whereas the $RM$ correlation
length was determined to be in the range of $4.5\ldots5.0$ kpc. These
ranges have to be considered as a systematic uncertainty since we are
not yet able to distinguish between these scenarios
observationally. Another systematic effect might be given by
uncertainties in the electron density itself. Varying the electron
density normalisation parameters ($n_{{\rm e1}}(0)$ and $n_{{\rm
e2}}(0)$) leads to a vertical displacement of the power spectrum while
keeping the same shape.

\begin{figure}[htb]
\resizebox{\hsize}{!}{\includegraphics{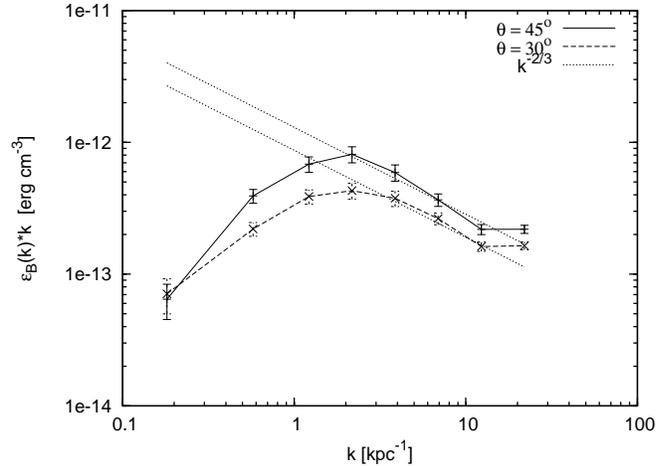}}
\caption[]{\label{fig:power_theta} Power spectra for two different
inclination angles $\theta = 30\degr$ and $\theta = 45\degr$ and an
$\alpha_B = 0.5$. For comparison a Kolmogorov-like power spectrum is
plotted as a straight dashed line. One can see that the calculated
power spectra follow such a power spectrum over at least one order of
magnitude. Note that the error bars are larger than in
Fig.~\ref{fig:power_alpha} because smaller bin sizes were used.}
\end{figure}

In order to study the influence of the inclination angle on the power
spectrum, we used an $\alpha_B = 0.5$, being in the middle of the most
likely region derived. For this calculation, we used smaller bins and
thus increased the number of bins to $n_l = 8$. We calculated the
power spectrum for two different inclination angles $\theta = 30\degr$
and $\theta = 45\degr$. The results are shown in
Fig.~\ref{fig:power_theta} in comparison with a Kolmogorov-like power
spectrum.

As can be seen from Fig.~\ref{fig:power_theta}, the power spectra
derived agree well with a Kolmogorov-like power spectrum over at
least one order of magnitude. For the inclination angle of $\theta =
30\degr$, we derived the following field and map properties \hbox{$B =
5.7 \pm 0.1 \, \mu$G}, \hbox{$\lambda_B = 3.1\pm0.3$ kpc} and
\hbox{$\lambda_{RM} = 6.7 \pm 0.7$ kpc}. For $\theta = 45\degr$, we
calculated \hbox{$B = 7.3 \pm 0.2 \, \mu$G}, \hbox{$\lambda_B =
2.8\pm0.2$ kpc} and \hbox{$\lambda_{RM} = 5.2 \pm 0.5$ kpc}. The value
of the log likelihood $\ln \mathcal{L}$ was determined to be slightly
higher for the inclination angle of $\theta = 30\degr$. The flattening
of the power spectra for large $k$s can be explained by small-scale
noise which we did not model separately.

Although the central magnetic field strength decreases with decreasing
scaling parameter $\alpha_B$, the volume-integrated magnetic field
energy $E_B$ within the cluster core radius $r_{{\rm c2}}$
increases. The volume-integrated magnetic field energy $E_B$ is
calculated as follows
\begin{equation}
E_B = 4 \pi \int _0 ^{r_{{\rm c2}}} \ud r\, r^2 \, \frac{B^2(r)}{8\pi}
= \frac{B_0^2}{2} \int _0 ^{r_{{\rm c2}}} \ud r \, r^2 \, \left(
\frac{n_{{\rm e}}(r)}{n_{{\rm e0}}} \right) ^{2 \alpha_B},
\end{equation}
where we integrate from the cluster centre to the core radius $r_{{\rm
c2}}$ of the second, non-cooling flow, component of the electron
density distribution.

\begin{figure}[htb]
\resizebox{\hsize}{!}{\includegraphics{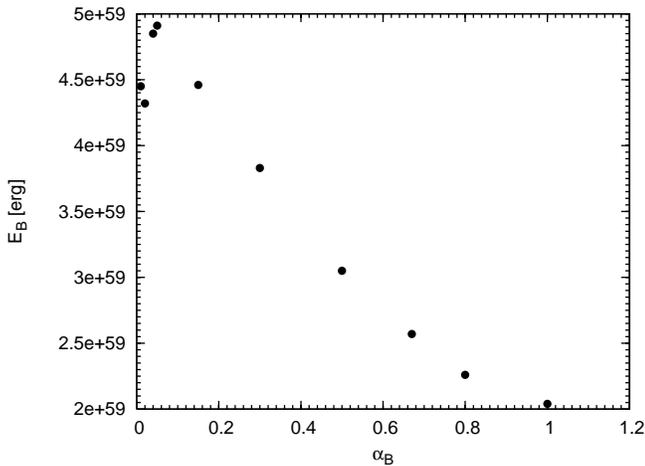}}
\caption[]{\label{fig:bradial} The integrated magnetic field energy
$E_B$ within the cluster core radius $r_{\rm c2}$ for the various
scaling parameters $\alpha_B$ also used in Fig.~\ref{fig:alpha_lnL} and
the corresponding central magnetic field strength $B_0$ as determined
by our maximum likelihood estimator.}
\end{figure}

We integrated the magnetic field profile for the various scaling
parameters and the corresponding field strengths which we determined by
our maximum likelihood estimator. The result is plotted in
Fig.~\ref{fig:bradial}. The higher magnetic energies for the smaller
scaling parameters which correspond to a lower central magnetic field
strength are due to the higher field strength in the outer parts of the
cool cluster core. This effect would be much more drastic if we had
extrapolated the scaling $B(r) \propto n_{\rm e}(r)^{\alpha_B}$ to
larger cluster radii and integrated over a larger volume.

\section{Conclusions\label{sec:conclusion}}
We presented a maximum likelihood estimator for the determination of
cluster magnetic field power spectra from $RM$ maps of extended
polarised radio sources. We introduced the covariance matrix for $RM$
under the assumption of statistically homogeneously-distributed
magnetic fields throughout the Faraday screen. We successfully tested
our approach on simulated $RM$ maps with known power spectra.

We applied our approach to the $RM$ map of the north lobe of Hydra
A. We calculated different power spectra for various window functions
being especially influenced by the scaling parameter between electron
density profile and global magnetic field distribution and the
inclination angle of the emission region. The scaling parameter
$\alpha_B$ was determined to be most likely in the range of
$0.1\ldots0.8$.

We realised that there is a systematic uncertainty in the values
calculated due to the uncertainty in the window parameter
itself. Taking this into account, we deduced for the central magnetic
field strength in the Hydra A cluster $B = (7 \pm 2)\,\mu$G and for the
magnetic field correlation length $\lambda_B = (3.0 \pm 0.5)$ kpc. If
the geometry uncertainties could be removed, the remaining statistical
errors are an order of magnitude smaller. The difference of these
values to the ones found in an earlier analysis of the same dataset of
Hydra~A which yielded $B = 12 \mu$G and \hbox{$\lambda_B = 1$ kpc}
\citep{2003A&A...412..373V} is a result of the improved $RM$ map using
the \textit{Pacman} algorithm \citep{2004astro.ph..1214D,
2004astro.ph..1216V} and a better knowledge of the magnetic cluster
profile, i.e. here $\alpha_B \approx 0.5$ \citep[instead of $\alpha_B =
1.0$ in ][]{2003A&A...412..373V}. However, the magnetic field strength
found in Hydra A supports the trend of relatively large magnetic fields
derived for cooling flow clusters from RM measurements reported in the
literature.

The cluster magnetic field power spectrum of Hydra A follows a
Kolmogorov-like power spectrum over at least one order of
magnitude. However, from our analysis it seems that there is a dominant
scale $\sim$ 3 kpc at which the magnetic power is concentrated.

\begin{acknowledgements}
We like to thank Greg Taylor for providing the polarisation data of
the radio source Hydra A and Klaus Dolag for the calculation of the $RM$
map using \textit{Pacman}. We like to thank Greg Taylor and Marat Gilfanov
for useful comments on the manuscript. 
\end{acknowledgements}

 
\bibliographystyle{aa}
\bibliography{cori}

\end{document}